\documentclass[12pt,a4paper]{article}
\pdfoutput=1
\usepackage{color}
\usepackage{amssymb,amsmath,bm,bbm}
\usepackage{epsf}
\usepackage{epsfig}
\usepackage{afterpage}
\usepackage{longtable}
\usepackage[dvipsnames]{xcolor}
\usepackage[linktoc=page,bookmarks=false,colorlinks=false,
linkbordercolor=RoyalBlue,citebordercolor=ForestGreen,urlbordercolor=CornflowerBlue]{hyperref}
\usepackage{latexsym,mathrsfs,dsfont}
\usepackage[normalem]{ulem} 
\usepackage[compress]{cite}
\usepackage{graphicx}
\usepackage{url}
\usepackage{paralist}
\usepackage{bbold}

\newcommand{\f}{\frac}
\newcommand{\ord}{\mathcal{O}}

\newcommand{\gev}{\, {\rm GeV}}
\newcommand{\mev}{\, {\rm MeV}}

\newcommand{\bsi}{B_6^{(1/2)}}
\newcommand{\bei}{B_8^{(3/2)}}

\def\epe{\varepsilon'/\varepsilon}
\newcommand{\beq}{\begin{equation}}
\newcommand{\eeq}{\end{equation}}
\newcommand{\be}{\begin{equation}}
\newcommand{\ee}{\end{equation}}
\newcommand{\bi}{\begin{itemize}}
\newcommand{\ei}{\end{itemize}}
\newcommand{\ba}{\begin{array}}
\newcommand{\ea}{\end{array}}
\newcommand{\beqa}{\begin{eqnarray}}
\newcommand{\eeqa}{\end{eqnarray}}
\newcommand{\bea}{\begin{eqnarray}}
\newcommand{\eea}{\end{eqnarray}}
\newcommand{\beqn}{\begin{eqnarray}}
\newcommand{\eeqn}{\end{eqnarray}}

\definecolor{red}{cmyk}{0,1,1,0.4}

\usepackage{fancyhdr}
\advance \headheight by 15.0truept       

\begin{document}


\vspace{-14mm}
\begin{flushright}
        {AJB-18-1}\\
CP3-18-21
\end{flushright}

\vspace{8mm}

\begin{center}
{\Large\bf
\boldmath{$K\to\pi\pi$ and $K-\pi$ Matrix Elements of the Chromomagnetic Operators from  Dual QCD}}
\\[12mm]
{\bf \large  Andrzej~J.~Buras ${}^a$ and Jean-Marc G\'erard${}^b$ \\[0.8cm]}
{\small
${}^a$TUM Institute for Advanced Study, Lichtenbergstr.~2a, D-85748 Garching, Germany\\
Physik Department, TU M\"unchen, James-Franck-Stra{\ss}e, D-85748 Garching, Germany\\[2mm]
${}^b$ Centre for Cosmology,
Particle Physics and Phenomenology (CP3), Universit{\'e} catholique de Louvain,
Chemin du Cyclotron 2,
B-1348 Louvain-la-Neuve, Belgium}
\end{center}

\vspace{8mm}

\abstract{%
\noindent
We perform for the {\it first time} a direct calculation of {\it on-shell} $K\to\pi\pi$ 
hadronic matrix elements of chromomagnetic operators (CMO) in the Standard Model and
beyond. To this end, we use the  successful Dual QCD (DQCD) approach in which we also 
consider {\it off-shell} $K-\pi$ matrix elements that allow the comparison with  
lattice QCD calculations presented recently by the ETM collaboration. Working in the SU(3) chiral limit, we find
for the single $B$ parameter $B_{\rm CMO}=0.33$.  Using the numerical results provided by the ETM collaboration we argue that only {\it small} corrections beyond that limit are to be expected. Our results are relevant for new physics scenarios in the context of  
the emerging $\epe$ anomaly strongly indicated within DQCD and 
supported by RBC-UKQCD lattice collaboration.
}

\setcounter{page}{0}
\thispagestyle{empty}
\newpage

\tableofcontents

\section{Introduction}
In view of the absence of direct signals of new physics (NP) from the LHC, 
indirect searches for it in the last years gained in importance. This is in particular
the case of flavour physics in $B$ meson and $K$ meson systems where 
several deviations from Standard Model (SM) expectations have been identified 
during the last years. While the so-called $B$ physics anomalies played 
 since 2013 the leading role in the indirect search for NP, also
the direct CP-violation in $K\to\pi\pi$ decays, represented by the ratio $\epe$,
 begins to play again  a  very important role in the tests of the Standard 
Model (SM) and more recently in the tests of its possible extensions. 
For recent reviews, see \cite{Buras:2013ooa,Buras:2016qia}. In fact 
there are strong hints for  sizable new physics contributions 
to $\epe$ from  Dual QCD  approach (DQCD)
\cite{Buras:2015xba,Buras:2016fys} that are supported to some extent by 
RBC-UKQCD lattice collaboration \cite{Bai:2015nea,Blum:2015ywa}. Recent
SM analyses at the NLO level can be found in  \cite{Buras:2015yba,Kitahara:2016nld} and a NNLO analysis is expected to appear soon \cite{Cerda-Sevilla:2016yzo}.
Most importantly, an improved result on $\epe$ from RBC-UKQCD lattice collaboration is expected this summer.

{This emerging $\epe$ anomaly} motivated several authors 
to look for various extensions of the SM which could bring the theory to agree with data as reviewed in \cite{Buras:2018wmb}. In most of the models the rescue comes from the
modification of the Wilson coefficient of the dominant electroweak LR penguin 
operator $Q_8$, but also solutions through a modified contribution of the dominant
QCD LR penguin operator $Q_6$ could be considered  \cite{Buras:2015jaq}. 

Here we want to address the issue of the role of 
chromomagnetic penguin operators in $\epe$ that attracted attention of several groups 
already two decades ago \cite{Bertolini:1994qk,Buras:1999da,Masiero:1999ub} and more recently in \cite{Bertolini:2012pu}.
 But the poor knowledge of their $K\to \pi\pi$ matrix elements prevented until recently a firm statement about their relevance. This changed through the first numerical lattice QCD value  \cite{Constantinou:2017sgv} of the related $K-\pi$
matrix element
which appears to be rather different from the one  found in the past in 
the chiral quark model \cite{Bertolini:1994qk}.

The main goal of the present paper is to study these matrix elements 
 in the framework of Dual QCD approach \cite{Bardeen:1986vp,Bardeen:1986uz,Bardeen:1986vz,Bardeen:1987vg,Buras:2014maa,Buras:2015xba,Buras:2016fys}.
  While 
 not as precise as ultimate lattice QCD calculations, 
this successful approximation to low-energy QCD offered over 
 many years an insight in the lattice results and often, like was the case
of the $\Delta I=1/2$ rule \cite{Bardeen:1986vz} 
and the parameter $\hat B_K$ \cite{Bardeen:1987vg}, provided results 
 almost three decades before this was possible with Lattice QCD. 
The agreement between results from DQCD  and Lattice QCD is
remarkable, in particular considering the simplicity of the former analytical approach 
 with respect to the very sophisticated and computationally demanding  numerical lattice QCD one. The most recent examples of this agreement are an explanation within DQCD of the pattern of values obtained by lattice QCD for the SM parameters
 $\bsi$ and $\bei$  entering $\epe$ \cite{Buras:2015xba,Buras:2016fys} and 
for the NP parameters $B_i$ entering $\varepsilon_K$ \cite{Buras:2018lgu}.

Our paper is organized as follows. The first two sections could be considered 
as a pedagogical introduction into DQCD which, we hope, will  make  
our discussion on chromomagnetic penguin operators clearer. In Section~\ref{sec:2} we recall few lessons gained from DQCD on the $\Delta I=1/2$ rule and in Section~\ref{sec:3} analogous lessons on QCD penguins.  In Section~\ref{sec:4} we present 
general formulae in DQCD from which hadronic matrix elements of chromomagnetic penguins for $K\to\pi\pi$ and  $K-\pi$ but also $K-K$ and $\pi-\pi$ can
be calculated. This we do in Section~\ref{sec:5} and 
compare our results with those 
obtained in the Chiral Quark Model and in lattice QCD. We summarize our results
in Section~\ref{sec:6}. Phenomenological applications of our results are left
for the future.

\boldmath
\section{A Brief Lesson from the $\Delta I=1/2$ Rule}\label{sec:2}
\unboldmath

  In the one-loop approximation, the first known impact of an asymptotically free theory for strong interactions on the $K\to\pi\pi$  weak decays, namely an octet enhancement \cite{Gaillard:1974hs,Altarelli:1974exa},  can easily be generalized to all electroweak processes in the large $N$ limit, $N$ being the number of colours. At the quark-gluon level, the effective operators run in the following way 
\be\label{QE1}
(J_L)^{ab}(J_L)^{cd}(\mu_{\rm SD})= (J_L)^{ab}(J_L)^{cd}(\mu)
-\left[\left(3\f{\alpha_s}{4\pi}\right)\ln(\f{\mu_{\rm SD}^2}{\mu^2})\right]\,(J_L)^{ad}(J_L)^{cb}(\mu)
\ee
with
\be\label{JL}
(J_{L})^{ab}=\bar q_L^b\gamma^\mu q_L^a, \qquad (a,b=u,\,d,\,s)
\ee
being any left-handed current with $q_{R,L}=(1\pm\gamma_5)/2$. In (\ref{QE1})
the $\mu_{\rm SD}$ and $\mu$ scales are assumed to be in the range of, say, 
$1\gev < \mu\le\mu_{\rm SD}<10\,\mu$  in order to safely apply perturbative QCD on the one hand, and elude standard leading-log resummation beyond the scope of our paper on the other hand. Further non-perturbative evolution below one GeV,  down to the factorization scale denoted here by $0$, is thus required to consistently estimate hadronic decay amplitudes. Such a one-loop evolution, involving the low-lying nonet of mesons only, turns out to already provide a rather consistent pattern \cite{Bardeen:1986vz}. At the hadronic level, we have indeed 
\be\label{ME1}
(J_L)^{ab}(J_L)^{cd}(M)= (J_L)^{ab}(J_L)^{cd}(0)-
 4\,\left(\frac{M}{4\pi f}\right)^2 (J_L)^{ad}(J_L)^{cb}(0)
\ee
with
\be\label{Jab}
(J_L)^{ab}(0)=i\f{f^2}{8}\left[\partial U U^\dagger-U\partial U^\dagger\right]^{ab},\qquad
U=\exp(i\sqrt{2}\frac{\pi}{f})
\ee
the left-handed pseudo-Goldstone currents in the chiral limit and $f\approx f_\pi=130.2\mev$. So, with our conventions (\ref{JL}) and (\ref{Jab}) for the quark and meson currents, the pseudo-Goldstone field and the associated left-handed electroweak currents transform as 
\be\label{UTR}
U\to g_L\, U\, g_R^\dagger,\qquad J_L\to g_L\, J_L\, g_L^\dagger
\ee
under the $\text{U(3)}_L\times\text{U(3)}_R$ acting
 separately on $q_L$ and $q_R$, respectively. This seemingly innocuous remark will turn out to be quite useful when fixing the chiral structure and relative signs of hadronic operators.

    From the generic mixing pattern (\ref{ME1}), one immediately concludes that the naive vacuum saturation approximation (VSA), namely
\be\label{VSA}
(J_L)^{ab}(J_L)^{cd}(\mu)= (J_L)^{ab}(J_L)^{cd}(0)+
 \f{1}{N} (J_L)^{ad}(J_L)^{cb}(0),\qquad {\rm (VSA)}
\ee
does not make sense. It is a scale-independent (stepwise) evolution down to zero momentum that completely misrepresents next-to-leading effects in a $1/N$ expansion based on the following dual counting:
\be\label{E6}
\f{\alpha_s}{4\pi}\propto \f{1}{N}\propto \frac{p^2}{(4\pi f)^2}\,.
\ee

As a matter of fact, strong interaction impact on electroweak processes do not suddenly die out around one GeV and the wrong-sign Fierz term in (\ref{VSA}) is only a part of the full (continuous) one-loop meson evolution (\ref{ME1}). Thus, as already emphasized long time ago \cite{Buras:1985yx,Bardeen:1986vp}, 
 the VSA gives theoretically consistent results only in the strict large
 N limit. Such is clearly the case for the $\Delta S = 2$ Fierz self-conjugate operator $J_L^{ds} J_L^{ds}$, as reviewed in \cite{Buras:2014maa}. The predicted $1/N$ flip of sign between (\ref{VSA}) and  (\ref{ME1}), discussed at length in \cite{Gerard:1990dx}
for the $\Delta S = 1$ Fierz conjugate operators  $J_L^{us} J_L^{du}$  and 
 $J_L^{ds} J_L^{uu}$, has been recently confirmed by Lattice QCD numerical simulations \cite{Boyle:2012ys}. 

   Our analytical Dual QCD approach provides a rather simple picture of the $\Delta I = 1/2$ rule far beyond its initial pulse (\ref{QE1}). The inclusion of the next-to-leading corrections to hadronic matrix elements can be viewed as the evolution of the operators from zero momentum to M. This fast meson evolution in 
(\ref{ME1}) is continued above $M$ as a slower quark-gluon evolution in (\ref{QE1}), the latter being improved by the usual QCD renormalization group equations applied on $\mu$ to reach the far-off Fermi scale $(\mu_{\rm SD}\approx M_W\gg 10\gev)$ in terms of Wilson coefficients. The link between $M$ and $\mu$ scales around one GeV is therefore subject to theoretical uncertainties. It has been demonstrated in  \cite{Buras:2014maa} that the inclusion of heavier mesons helps turning the quadratic evolution in (\ref{ME1}) into the logarithmic one in (\ref{QE1}). However, in our paper, we are not concerned with this question since we only focus on mixing patterns to identify penguin-like operators deliberately overlooked in (\ref{QE1}) and (\ref{ME1}) until now for the sake of the forthcoming developments.

\section{Strong Penguin Operators}\label{sec:3}
\subsection{Short-distance mixing pattern}

   The existence of so-called penguin operators \cite{Shifman:1975tn} in the short-distance (SD) evolution of the $\Delta S = 1$ weak operators below the charm mass scale can also be generalized in the large $N$ limit \cite{Bardeen:1986uz}. As a consequence, we have the following substitution for the Fierz operator emerging from $J_L^{ab}J_L^{cd}$ in 
(\ref{QE1})
\be\label{S1}
J_L^{ad}J_L^{cb}\Rightarrow J_L^{ad}J_L^{cb}+\f{1}{9}Q_P^{ad,cb}\,,
\ee
where
\be\label{QP}
Q_P^{ad,cb}=\delta^{ad}(J_L^{cq}J_L^{qb}-2 D_L^{cq}D_R^{qb})+(ad\leftrightarrow cb)\,,
\ee
with a summation over repeated flavour indices $q$ understood hereafter, and
\be\label{DRL}
D_{R,L}^{ab}=\bar q^b_{L,R} q^a_{R,L}\,
\ee
any chiral quark density operator.      

For $(ad,cb)=(uu,ds)$, the result in (\ref{S1}) can be directly obtained from the relevant entries
in  the anomalous dimension matrix describing the mixing of the current-current 
operator $Q_2$ with the QCD penguin operators $Q_{4,6}$ and the mixing 
of $Q_2$ with the second current-current operator $Q_1$ in \cite{Buras:1998raa}
\be
\frac{\gamma_{24}^{(0)}}{\gamma_{21}^{(0)}} =\frac{\gamma_{26}^{(0)}}{\gamma_{21}^{(0)}}=\frac{1}{9} \,.
\ee

\subsection{Long-distance mixing pattern}     
We find it quite remarkable that the existence of these penguin operators could have been anticipated from a simple long-distance (LD) evolution, well before the advent of  $\text{SU}(N)$ - QCD. Indeed, working again with the low-lying pseudoscalars, we also generate additional nonet operators in the one-loop meson evolution down to the factorization scale. As demonstrated in \cite{Fatelo:1994qh} such an evolution amounts to the following substitution for the Fierz operator emerging from $J_L^{ab}J_L^{cd}$
 in (\ref{ME1}):
\be\label{ME2}
J_L^{ad}J_L^{cb}\Rightarrow J_L^{ad}J_L^{cb}-\f{1}{2}\left[\delta^{ad}(J_LJ_L)^{cb}+(ad\leftrightarrow cb)\right]\,.
\ee

Moreover, the addition of the light vector mesons in the loop transmutes partially the quadratic dependence on $M$ into a logarithmic one as expected from a full dual picture of QCD, namely with an infinite number of mesons, but does not modify this mixing pattern in the chiral limit \cite{Buras:2014maa}. The reason why such is the case is rather simple to understand in our dual QCD approach: the operators 
$J_L^{ab} \text{Tr}(J_L)$ have no LD evolution. Indeed, for these operators, non-factorizable meson exchanges between the two currents are forbidden since the trace of the left-handed current (\ref{Jab}) is just proportional to the derivative of a single flavour-singlet state, i.e., $\text{Tr}(J_L)=-\sqrt{3/2}\partial \eta^0$.

\boldmath
\subsection{Matching SD and LD patterns around one $\gev$}
\unboldmath
The inferred pattern for the LD meson evolution most easily derived with the help of a background field method developed in  \cite{Fatelo:1994qh} allows thus a consistent matching of the meson and quark-gluon evolutions if
\be\label{QPF}
Q_P^{ad,cb} =-\f{9}{2}\left[\delta^{ad}(J_LJ_L)^{cb}+(ad\leftrightarrow cb)\right]\,,
\ee
as obtained by equating (\ref{S1}) and (\ref{ME2}). Having (\ref{QPF}) and 
(\ref{QP}) at our disposal,  we can even disentangle the density-density component of $Q_P$  in (\ref{QP}) from its current-current one, i.e., 
\be\label{DLDR}
D_L^{cq}D_R^{qb}\approx + \f{11}{4}J_L^{cq}J_L^{qb}
\ee
around the optimal matching scale to be consistently estimated here below. This disentanglement is obviously impossible to achieve by means of a pure Chiral Perturbation Theory in which any short-distance information from perturbative QCD is fully encoded in low energy parameters to be determined through physical observables. The same comment will evidently apply to what follows.

\section{Chromomagnetic Penguin Operators in and Beyond the Standard Model}\label{sec:4}
Contrary to the current-current or density-density operators considered until now, the two-quark chromomagnetic operators cannot be factorized and a parametrization of their hadronic matrix elements with respect to a naive VSA is thus impossible. Yet, they can be identified in the large $N$ limit through their mixing with the former ones. So, for that purpose we proceed with the same methodology as in the anatomy of the penguin operators (\ref{QP}).

\subsection{Short-distance mixing pattern}
The mass-dependent {\it dimension-six} chromomagnetic operators defined by 
\be\label{QG0}
Q_G^{ad,cb}\equiv \f{g_s}{16\pi^2}\left\{\delta^{ad}\,[m_c(\bar q^b_L\sigma^{\mu\nu}G_{\mu\nu}q^c_R)+m_b(\bar q^b_R\sigma^{\mu\nu}G_{\mu\nu}q^c_L]+(ad\leftrightarrow cb)\right\},
\ee
with $G_{\mu\nu}=t^aG^a_{\mu\nu}$, are induced from the current-currents ones by considering one-loop Feynman diagrams calculated with massless quarks except on external lines where a single 
$R-L$ chirality flip is required \cite{Ciuchini:1993fk,Buras:1998raa}\footnote{These diagrams are most familiar from the studies of $b\to s$\,gluon transition, but here of course $m_c$ and $m_b$ stand  for appropriate light quark masses only.}. In the large $N$ limit, the substitution (\ref{S1}) for the Fierz operator in (\ref{QE1}) becomes then
\be\label{CH}
J_L^{ad}J_L^{cb}\Rightarrow J_L^{ad}J_L^{cb}+\f{1}{9}Q_P^{ad,cb}+\f{11}{54}\,N\, Q_G^{ad,cb}\,.
\ee
For $(ad,cb)=(uu,ds)$, this can again be obtained by  looking up the anomalous dimension entry 
 $\gamma^{(0)}_{28}$ in  \cite{Ciuchini:1993fk,Buras:1998raa} and comparing it 
to  $\gamma^{(0)}_{21}$. To this end, the non-leading in $N$ term should be dropped.
\subsection{Long-distance mixing pattern}
To guarantee the same chirality flip at the meson level, we have to introduce a linear dependence on the light quark mass matrix $m =\text{diag.}(m_u, m_d, m_s)$ for the corresponding hadronic currents. In the large $N$ limit, such dependence turns out to be uniquely defined \cite{Bardeen:1986vz}. The transformation laws (\ref{UTR}) supplemented with $m\to g_L m g_R^\dagger$ determine the chiral structure of these improved currents:  
\be\label{VAc}
J_L^{ab} =i\frac{f^2}{8}\left[(\partial U)U^\dagger-U(\partial U^\dagger)+
\frac{r}{\Lambda^2_\chi} [(\partial U)m^\dagger-m(\partial U^\dagger)]\right]^{ab}
,
\ee
with $\Lambda_\chi$ and $r$, two scales to be extracted from the SU(3) splitting among the weak decay constants and the (running) quark masses. In particular, for the charged kaon and pion we have 
\be\label{FKFP}
f_K=f(1+\f{m_K^2}{\Lambda^2_\chi}),\qquad f_\pi=f(1+\f{m_\pi^2}{\Lambda^2_\chi})
\ee
and
\be\label{rr}
m_K^2=\f{r}{2}(m_s+m_d),\qquad m_\pi^2=\f{r}{2}(m_u+m_d)
\ee
implying
\be\label{input}
\Lambda_\chi\approx 1.09\gev, \qquad r(1\gev)\approx 3.75\gev
\ee
from
$f_K/f_\pi=(1.193\pm0.003)$ and $(m_s+m_d)(1\gev)=132\mev$, respectively \cite{Aoki:2016frl}. Compared to large $N$ Chiral Perturbation Theory, there is one-to-one correspondence with the low energy parameters introduced in \cite{Gasser:1983yg,Gasser:1984gg}
\be
\Lambda^2_\chi=\frac{f^2}{8L_5}, \qquad r=2B_0\,.
\ee

     Let us underline that the transformation laws (\ref{UTR}) together with the inputs (\ref{input}) fix once for all the (positive) sign of the chiral correction to the left-handed current in (\ref{VAc}) as well as the (negative) sign of the chiral correction to the left-handed and right-handed densities:
\be\label{RLd}
D_{L(R)}^{ab}=-\frac{f^2}{8}r\left[U^{(\dagger)}-\frac{1}{\Lambda_\chi^2}\partial^2U^{(\dagger)}\right]^{ab}.
\ee

Consequently, at the meson level the density-density component of the strong penguin operators (\ref{QP}) reads
\be
D_L^{cq}D_R^{qb} = + \f{r^2}{2\Lambda_\chi^2} J_L^{cq}J_L^{qb}+\ord(\f{1}{\Lambda_\chi^4})
\ee
and allows us to estimate now the optimal matching scale $\mu_M$ between the SD and LD evolutions on the basis of (\ref{DLDR}).
 The relation  
\be
r(\mu_M)=\sqrt{\f{11}{2}}\, \Lambda_\chi\, \approx 2.55\gev,
\ee
comforts us in the idea that the Dual QCD approach is quite consistent around 
$0.7\gev$.  

Using once more the background field method for the meson evolution  \cite{Fatelo:1994qh}, 
at $\ord(1/\Lambda_\chi^2)$ we also get an additional operator such that the substitution (\ref{ME2}) for the Fierz operator in (\ref{ME1}) becomes
\be\label{ME3}
J_L^{ad}J_L^{cb}\Rightarrow J_L^{ad}J_L^{cb}-\f{1}{2}[\Delta^{ad,cb}+\Delta^{cb,ad}]
\ee
with 
\be
\Delta^{ad,cb}= \delta^{ad}\left[(J_LJ_L)+\frac{r}{\Lambda_\chi^2}(mU^\dagger J_L J_L+J_L J_L Um^\dagger)\right]^{cb}\,.
\ee
 
 An induced tadpole 
\be
T_G^{ad,cb}\propto [(mU^\dagger+Um^\dagger)^{ad}\delta^{cb}+(ad\leftrightarrow cb)]
\ee
has been ignored in (\ref{ME3}) since the $\langle \pi\pi|T_G|K^0\rangle $ 
matrix element for the on-shell $K^0\to \pi\pi$  decay amplitude is exactly cancelled by a pole contribution involving the strong $K^0\to \pi\pi \bar K^0$
vertex followed by the $\bar K^0$  annihilation in the vacuum through the non-zero  $\langle \bar K^0|T_G| 0\rangle $ weak matrix element 
\cite{Donoghue:1985ae}.  Were we considering unphysical (off-shell) hadronic matrix elements, such would not be allowed, in principle.

\boldmath
\subsection{Matching SD and LD patterns around one $\gev$}
\unboldmath    
Here, there is no ambiguity in identifying the single {\it dimension-six} chromomagnetic operator $Q_G$ at low momenta. By comparing  the last term in (\ref{CH}) with $N = 3$ to
(\ref{ME3}), we obtain 
\be\label{QG1}
Q_G^{ad,cb} =-\frac{9}{11}\frac{r}{\Lambda_\chi^2}[\delta^{ad}(mU^\dagger J_L J_L+J_L J_L Um^\dagger)^{cb}+ (ad\leftrightarrow cb)]
\ee
which appears to be similar to $Q_P$ in (\ref{QPF}), both in the chiral structure and sign.

 Mass - independent {\it dimension-five} chromomagnetic operators
\be
Q_g^{ad,cb}(\pm)\equiv \f{g_s}{16\pi^2}\left\{\delta^{ad}\left[(\bar q_L^b\sigma^{\mu\nu}G_{\mu\nu}q^c_R)\pm (\bar q_R^b\sigma^{\mu\nu}G_{\mu\nu}q^c_L)\right]+(ad\leftrightarrow cb)\right\}
\ee
transforming like $(U^\dagger \pm U)$  under chiral symmetry may arise beyond the Standard Model, and in particular in SUSY extensions \cite{Buras:1999da,Masiero:1999ub}. From (\ref{QG1}), we directly infer that in the large $N$ limit such an SU(3) chiral operator would hadronize in the following way:
\be\label{QG7}
Q_g^{ad,cb}(\pm)=-\f{9}{11}\frac{r}{\Lambda_\chi^2}[\delta^{ad}(U^\dagger J_L J_L\pm J_L J_L U )^{cb}+ (ad\leftrightarrow cb)]\,
\ee
around the matching scale $\mu_M$. This is consistent with the fact 
that  for $(ad,cb)=(uu,ds)$ the entry  $\gamma^{(0)}_{28}$ in  \cite{Ciuchini:1993fk,Buras:1998raa} 
will not be modified by replacing $Q_G$ by $Q_g$ so that the numerical factor $9/11$ will remain unchanged.

\boldmath
\section{The Case of $\Delta S = 1$ Matrix Elements}\label{sec:5}
\unboldmath
\subsection{Dual QCD}
Driven by the $\Delta S=1$ operator $J_L^{us}J_L^{du}$, only the first term proportional to the $\delta^{uu}$  Wick contraction
in (\ref{QG1}) survives and             
\be\label{QG2}
Q_G^{uu,ds}=-\frac{9}{176}\left(\frac{f_\pi^4r}{\Lambda_\chi^2}\right)[mU^\dagger \partial U\partial U^\dagger
+\partial U\partial U^\dagger U m^\dagger]^{ds}\,,
\ee
meaning that the unwanted internal up quark mass contribution is automatically removed as it should. As we will prove in Section~\ref{LQCDS}, such is not the case for $\Delta S=0$ operators for which a second Wick contraction is possible, a crucial feature for our interpretation of the SU(3) chiral limit on lattice QCD.
The hadronic matrix elements of this dimension-six chromomagnetic operator for the parity-violating $K^0\to\pi\pi$ processes $(\pi\pi=\pi^+\pi^-,\pi^0\pi^0)$: 
\be\label{Parity}
\langle \pi\pi|Q_G^{uu,ds}|K^0\rangle=-i\left[\f{9}{22}\right]\frac{m_\pi^2}{\Lambda_\chi^2} f_\pi (m_K^2-m_\pi^2)
\ee
are suppressed by almost two orders of magnitude and of opposite sign with respect to the Standard Model current-current one:  
\be
\langle \pi^+\pi^-|J_L^{us}J_L^{du}|K^0\rangle=\langle \pi\pi|(J_LJ_L)^{ds}|K^0\rangle= +i\frac{f_\pi}{4}
(m_K^2-m_\pi^2)\,.
\ee
With its opposite sign Wilson coefficient compared to the $J_LJ_L$ one, as already sensed from (\ref{QPF}) and (\ref{CH}), the chromomagnetic operator $Q_G$ cannot help much for the $\Delta I = 1/2$ rule and $\epe$.

      Similarly, for the dimension-five chromomagnetic operators beyond the Standard Model
\be\label{QG3}
Q_g^{uu,ds}(\pm)=-\frac{9}{176}\left(\frac{f_\pi^4r}{\Lambda_\chi^2}\right)[U^\dagger \partial U\partial U^\dagger
\pm \partial U\partial U^\dagger U)]^{ds}\,,
\ee
we obtain the following matrix elements 
\be\label{Qg-}
\langle \pi\pi|Q_g^{uu,ds}(-)|K^0\rangle=+i\left[\f{9}{22}\right]\frac{m_\pi^2}{\Lambda_\chi^2} f_\pi r(\mu)
\ee
  
\be\label{Qg+}
\langle \pi^+|Q_g^{uu,ds}(+)|K^+\rangle=- \left[\frac{9}{22}\right]\frac{p_K\cdot p_\pi}{\Lambda_\chi^2} f^2_\pi r(\mu)
\ee
with Wilson coefficients $C^\pm_g$ that depend on the UV completion of the 
theory.

Recalling the parametrization of $K^0\to\pi^+\pi^-$ and $K^0\to\pi^0\pi^0$
in terms of isospin amplitudes $A_{0,2}$ that enter the evaluation of $\epe$ and the $\Delta I=1/2$ rule
\begin{equation}\label{ISO2}
A(K^0\rightarrow\pi^+\pi^-)=\left[A_0 e^{i\delta_0}+ \sqrt{\frac{1}{2}}
 A_2 e^{i\delta_2}\right]~,
\end{equation}
\begin{equation}\label{ISO3}
A(K^0\rightarrow\pi^0\pi^0)=\left[A_0 e^{i\delta_0}-\sqrt{2} A_2 e^{i\delta_2}\right]\,,
\end{equation} 
we find the chromomagnetic penguin contributions to $A_{0,2}$ as follows
\be\label{BG1}
(\Delta A_0)_{CMO}= C_g^-(\mu)\,\left[\f{9}{11}\right]\frac{m_\pi^2}{\Lambda_\chi^2} f_\pi \frac{m_K^2}{m_s(\mu)+m_d(\mu)}, \qquad A_2=0\,,   \qquad ({\rm DQCD}).
\ee
Consequently 
\be\label{BG2}
(\Delta A_0)_{CMO} = 4.1\, C_g^-(\mu) \,\left[\frac{100\mev}{m_s(\mu)+m_d(\mu)}\right] \, 10^{-3} \gev^2, \qquad ({\rm DQCD}).
\ee
Note that dimension of $C_g^-(\mu)$ is $\gev^{-1}$.

These  first direct calculations to date of chromomagnetic contributions to  physical decay amplitudes in  the context of a successful approximation to low-energy QCD 
constitute the main results of our paper.
In what follows we will compare our 
results with the ones obtained in Chiral Quark Model long time ago \cite{Bertolini:1994qk} and in lattice QCD quite recently \cite{Constantinou:2017sgv}, where 
  in both cases $K-\pi$ transitions have been considered.

\subsection{Chiral Quark Model}
  As a matter of fact, the $m^2_\pi$  kinematical suppression displayed in (\ref{Parity}) for the hadronic matrix element of the dimension-six chromomagnetic operator has been first highlighted within a specific Chiral Quark Model (CQM) \cite{Bertolini:1994qk}. In this paper, the so-called magnetic dipole operator $(Q_{11}$) is twice our chromomagnetic one ($Q_G$) defined in (\ref{QG0}). Consequently, after bosonization and normalization with respect to the parity-even off-shell ($K^+- \pi^+$) hadronic matrix element (with the tadpole contribution simply neglected), the authors of  \cite{Bertolini:1994qk} eventually obtain the following expression  

\be\label{QG4}
Q_G^{uu,ds}=+\frac{11}{32}\left[\frac{f_\pi^4r}{(4\pi f_\pi)^2}\right]
[m^\dagger U  \partial U^\dagger\partial U
+\partial U^\dagger\partial U U^\dagger m]^{ds}\, \qquad {\rm (CQM)}
\ee
if, on the basis of (\ref{RLd}), we consistently substitute our $r(\mu)$ for their $(\bar q q)$ quark condensate. 

At first sight, (\ref{QG4}) and (\ref{QG2}) seem to disagree both in the chiral structure and sign. However, in  (\ref{QG4}), the pseudo-Goldstone fied $U$ transforms into $g_R U g_L^\dagger$ under the chiral $\text{U(3)}_L\times\text{U(3)}_R$ symmetry, i.e. as our $U^\dagger$ according to (\ref{UTR}), to match with the definition (\ref{QG0}) for the chromomagnetic operator at the quark-gluon level. The way $U$ (or $U^\dagger$) initially transforms under the chiral symmetry is of course just a matter of convention. It is only once parity-violating transitions such as 
$K^+(\pi^+)\to\mu^+\nu_\mu$ or $K\to\pi\pi$  are also considered that this freedom is definitely lost. Indeed, $U(\pi)$ turns into $U^\dagger(\pi) = U(-\pi)$ under parity. As a consequence the sign of all the parity-odd or, equivalently, of all the parity-even hadronic matrix elements derived from (\ref{QG2}) has to be flipped before any comparison with the
ones derived on the basis of (\ref{QG4}) can be made. For completeness, we should also mention that in \cite{Bertolini:2012pu} the factor (11/32) in (\ref{QG4}) has been corrected to (1/4). However, when comparing with lattice results 
below we do not introduce this change in order to agree with the definition 
of the forthcoming parameter $B_{\rm CMO}$ used in \cite{Constantinou:2017sgv}.

\subsection{Lattice QCD}\label{LQCDS}
 Obviously inspired by the CQM pathfinder result (\ref{QG4}), the authors of 
\cite{Constantinou:2017sgv} choose to normalize the $\Delta S = 1$ dimension-five chromomagnetic operator (CMO) within the same chiral convention:
\be\label{QG6}
Q_g^{uu,ds}(\pm)\equiv +\frac{11}{256\pi^2}\left[\frac{f_\pi^2m_K^2}{m_s+m_d}\right]B_{\rm CMO}
[U  \partial U^\dagger\partial U
\pm \partial U^\dagger\partial U U^\dagger]^{ds}\, \quad {\rm (Lattice~QCD)}\,,
\ee
where we have used (\ref{rr}) to express $r$ in terms of $m_K$ and $m_{s,d}$ 
as done in \cite{Constantinou:2017sgv}. These authors
also focus first on the parity-even $\langle \pi^+|Q_g^{uu,ds}(+) |K^+\rangle$ hadronic matrix element (with the tadpole contribution quite carefully substracted in the renormalized operator on a lattice) to obtain a value smaller than the CQM one for the corresponding B-parameter, i.e.,  
\be\label{CMO1}
B_{\rm CMO}^{K\pi} = 0.273(69)\, \qquad      {\rm (Lattice~QCD)}     
\ee                                       
at the physical pion and kaon point. In Dual QCD, the hadronic matrix elements given in  (\ref{Qg-}) and (\ref{Qg+}) translate, after the conventional flip of sign for the second one, into a {\it single} $B$-parameter comparable in size to the Lattice QCD one:  
\be\label{CMO2}
B_{\rm CMO} =\left(\f{18}{121}\right)\frac{3}{N}
\left[\f{4\pi f_\pi}{\Lambda_\chi}\right]^2\approx 0.33 \qquad {\rm (Dual~QCD)}.
\ee
Here, the colour factor $N$ present in (\ref{CH}) is re-introduced to exhibit the agreement with the $1/N$ expansion rules (\ref{E6}).

   On the one hand, a direct confrontation of (\ref{CMO1}) and (\ref{CMO2}) suggests relatively small chiral corrections despite the unexpected  $m^2_\pi$ kinematical suppression observed in (\ref{Qg-}) for the {\it physical}
$K^0\to\pi\pi$ transition. On the other hand, new lattice results on {\it unphysical}  $K-K$ and $\pi-\pi$ hadronic matrix elements in the SU(3) chiral limit give \cite{Constantinou:2017sgv} 
\be\label{34a}
 B^{(\Delta S=1)}_{\rm CMO}=0.076\,(23)\,,\qquad ({\rm Lattice~QCD}).
\ee
According to the authors of \cite{Constantinou:2017sgv}, their numerical results in (\ref{CMO1}) and (\ref{34a}) imply 
a strong enhancement factor equal to 4 possibly via $\ord[(m_K-m_\pi)^2/(4\pi f_\pi)^2]$ corrections  in contradiction to our expectations from the 
comparison of (\ref{CMO1}) and (\ref{CMO2}).

Here we would like to present our {\it own} interpretation of the numerical 
studies in \cite{Constantinou:2017sgv} as seen from the point of view of 
SU(3) chiral symmetry. To this end it should be emphasized 
that the result in (\ref{34a})  has 
been obtained  by considering the matrix elements 
\be\label{ETM}
\langle K|\hat O_{CM}|K\rangle \equiv [\langle K|\hat O_{CM}|\pi \rangle]_{m_d=m_s},\quad  \langle \pi|\hat O_{CM}|\pi\rangle \equiv [\langle K|\hat O_{CM}|\pi \rangle]_{m_s=m_d=m_{ud}}
\ee
with {\it unphysical} $K(\pi)$ in the initial (final) state respectively since 
$\hat O_{CM}$  is still the $\Delta S=1$ renormalized $g_s\bar s\sigma^{\mu\nu}G_{\mu\nu} d$ operator on a lattice.

However,  interpreting the ETM's results (\ref{CMO1}) and (\ref{34a}) in terms of a single parameter $B_{\rm CMO}$ defined in (\ref{QG6}) requires to consider the physical $K$ and $\pi$ (pseudo-Goldstone) states inserted by definition 
in the $U$ field in (\ref{Jab})
 and, 
consequently, $\Delta S=0$ rather than $\Delta S=1$ operators. In such an SU(3) chiral extrapolation the second term in (\ref{QG7})
which is, as stressed at the beginning of Section~\ref{sec:5}, absent in 
$\Delta S=1$ transitions contributes a factor of two already at the operator level to $\Delta S=0$ transitions. Indeed this factor
comes from the fact that contrary to what happens for the $(K^+- \pi^+)$ transition, the external up quark mass also contributes in  the strange or down  quark one-loop Feynman diagrams responsible for the $(K^+- K^+)$ or $(\pi^+- \pi^+)$ one, namely
\be
[s-(uu)-d]\,,\quad   \text{for} \quad (\Delta S=+1),
\ee
and
\be
[s-(uu)-s] + [u-(ss)-u]\,\quad   \text{or} \quad   [d-(uu)-d] + [u-(dd)-u]   \quad   \text{for} \quad     (\Delta S=0)\,.
\ee

In other words, only the 
$\delta^{ad}$  Wick contraction survives in (\ref{QG1}) and consequently in (\ref{QG7}) for the $Q_g^{uu,ds}$ operator while both $\delta^{ad}$ {\it and} $\delta^{cb}$ Wick contractions contribute  for the SU(3)-related $Q_g^{uu,ss}$ and $Q_g^{uu,dd}$ operators.

But this is not the whole story. When calculating loop induced CMO hadronic matrix elements, one also has to take into account that the primary Hamiltonian is hermitian. In our DQCD approach, we have
\be
J_L^{us} J_L^{du}+J_L^{su} J_L^{ud}\qquad \text{for}\qquad (\Delta S=\pm 1)
\ee
and 
\be
J_L^{us} J_L^{su}+J_L^{su} J_L^{us},\quad \text{or}\quad  J_L^{ud} J_L^{du}+J_L^{du} J_L^{ud},\quad \text{for}\quad (\Delta S=0)\,.
\ee
Therefore, in the SU(3) chiral limit another factor of two distinguishes the 
emerging $\Delta S=0$ CMO quantum transitions from the $\Delta S=+1$ ({\it or}  $\Delta S=-1$) one for {\it physical} $K^+$ and $\pi^+$ states.

In DQCD, the unphysical $m_d=m_s$ limit would imply $m_{\pi^+}=m_{K^+}$ 
and $Q_g^{uu,ds}=Q_g^{uu,ss}=Q_g^{uu,dd}$.
On the basis of the Wick's theorem and unitarity in S-matrix expansion, 
we have thus argued that the result in (\ref{34a}) should also be multiplied by 4 and  turned into
\be\label{34b}
 B_{\rm CMO}=0.304\,(92)\,
\ee
before any quantitative comparison  with the rather consistent results displayed
in  (\ref{CMO1}) and (\ref{CMO2}) is made. An explicit lattice QCD calculation of the $K-K$ and $\pi-\pi$ matrix elements with physical initial and final 
states would of course be useful to confirm our claim.

 In view of the results  (\ref{CMO1}),
 (\ref{CMO2}) and (\ref{34b}) it is justified to expect only small chiral corrections to DQCD results for the contribution of chromomagnetic penguin 
operator to the $K\to\pi\pi$ isospin amplitude in (\ref{BG2}).  This is 
an important finding as the presence of very large chiral corrections to
 the only existing on-shell calculation of 
$K\to\pi\pi$ hadronic matrix element presented here would have diminished its importance in  the phenomenological 
studies of $\epe$ and $\Delta I=1/2$ rule.

All these results can be consistently derived from the full SU(3) chiral operator 
\be\label{QG11}
Q_g^{ad,cb}(\pm)=\frac{11}{256\pi^2}\left[\frac{f_\pi^2m_K^2}{m_s+m_d}\right]B_{\rm CMO}
\left[(U  \partial U^\dagger\partial U
\pm \partial U^\dagger\partial U U^\dagger)^{cb}\delta^{ad}+(ad\leftrightarrow cb)\right]
\ee
which generalizes (\ref{QG6}) and is valid for both $\Delta S=\pm 1$ 
{\it and }
$\Delta S=0$ transitions involving physical $K$ and $\pi$ states.

\section{Summary and Conclusion}\label{sec:6}

        We have considered the successful Dual QCD approach developed in the 1980s to estimate the impact of chromomagnetic operators (CMO) acting either in the Standard Model or beyond. We would like to emphasize that this is {the}
first direct calculation of the physical $K\to\pi\pi$ 
hadronic matrix elements {of} these operators to date as until now only matrix
elements for the transition $K-\pi$ have been calculated.

Our main results are as follows
\begin{itemize}
\item
In the Standard Model, the $\Delta S = 1$  hadronic matrix element of the dimension-six operator $Q_G$ proportional to $(m_s-m_d) [\bar s \sigma^{\mu\nu} G_{\mu\nu}\gamma^5d]$ is found to be kinematically suppressed by almost two orders of magnitude and of opposite sign with respect to the current-current one at the origin of the
$ \Delta I = 1/2$ rule: 
\be
\frac{\langle \pi\pi| Q_G|K^0\rangle}{\langle \pi\pi|J_L J_L|K^0\rangle}=
-\f{18}{11}(\f{f_K}{f_\pi}-1) \frac{m_\pi^2}{m_K^2-m_\pi^2} \approx -0.03\, .
\ee
\item
Beyond the Standard Model, possible dimension-five $Q_g (\pm)$ operators proportional to $[\bar s \sigma^{\mu\nu} G_{\mu\nu}(1,\gamma^5)d]$
are normalized with respect to a previous Chiral Quark Model calculation such that
\be\label{FINAL}
B_{\rm CMO} =\left(\f{288\pi^2}{121}\right)
(\f{f_K}{f_\pi}-1) \f{f_\pi^2}{m_K^2-m_\pi^2} 
\approx 0.33 \,.
\ee
 As a consequence of  our interpretation of recent lattice QCD results on $K-\pi$, $K-K$, $\pi-\pi$ transitions in terms of the full $SU(3)$ chiral operator (\ref{QG11}), we do not expect 
 sizeable chiral corrections for the B-parameter (\ref{FINAL}) associated with the on-shell $K^0\to\pi\pi$ transition. Phenomenological analyses with $B_{\rm CMO}\approx 1-4$  are therefore definitely ruled out.  
\end{itemize}

In summary the main result of our paper, to be used in phenomenological analyses
of the $\Delta I=1/2$ rule and of the ratio $\epe$, is the chromomagnetic 
penguin contribution to the isospin amplitude $A_0$ which reads
\be\label{BG5}
(\Delta A_0)_{CMO}\approx  4\, C_g^-(2\gev) \, \, 10^{-3} \gev^2\, .
\ee

    In the past, our Dual QCD approach has proved  to be a powerful analytical tool to either anticipate lattice QCD results on Standard Model hadronic matrix elements (e.g., with $B_K = 3/4$ in \cite{Buras:1985yx}) or shed new light on them (e.g., with 
$\bsi<\bei < 1$  in \cite{Buras:2015xba}. Recently also an insight in the values of NP $B_i$ parameters entering $\varepsilon_K$ and obtained by lattice QCD could be gained in this manner \cite{Buras:2018lgu}.

In this paper, we predict $B_{\rm CMO}\approx 1/3$ for the {\it on-shell} $K^0\to\pi\pi$  transitions induced by chromomagnetic operators that are likely to arise in physics beyond the Standard Model. While the recent calculation of {\it off-shell} $K-\pi$ matrix elements of chromomagnetic penguin operators (CMO) in 
\cite{Constantinou:2017sgv} 
should be considered as an important progress, 
we are looking forward to a 
direct calculation of the {\it on-shell} $K\to\pi\pi$ hadronic matrix elements of these operators  by the QCD lattice community or another analytical method that provides consistent matching between LD and SD QCD dynamics.

\section*{Acknowledgements}
We thank Christoph Bobeth, Vittorio Lubicz, Fabio Maltoni and Silvano Simula for discussions.
This research was supported by the DFG cluster
of excellence ``Origin and Structure of the Universe''.

\renewcommand{\refname}{R\lowercase{eferences}}

\addcontentsline{toc}{section}{References}

\bibliographystyle{JHEP}
\bibliography{Bookallrefs}
\end{document}